\documentstyle[11pt]{article}

\renewcommand{\theequation}{\arabic{section}.\arabic{equation}}

\begin{document}
\baselineskip 19pt
\parskip 7pt
\thispagestyle{empty}


\vspace{24pt}

\begin{center}
{\large\bf 
New integrable extension of the Hubbard chain with 
variable range hopping
}

\vspace{24pt}

Shuichi {\sc Murakami}\ 
\footnote[3]{E-mail:{\tt murakami@appi.t.u-tokyo.ac.jp}} 

\vspace{8pt}
{\sl Department 
of Applied Physics, Faculty of Engineering,} \\
{\sl University of Tokyo,} \\
{\sl Hongo 7-3-1, Bunkyo-ku, Tokyo 113-8656, Japan.} 

(Received:\makebox[15em]{} 

\end{center}

\vspace{6pt}

\begin{center}
{\bf Abstract}
\end{center}

New integrable variant of the one-dimensional Hubbard model with 
variable-range correlated hopping is studied.
The Hamiltonian is constructed by applying
the quantum inverse scattering method on the infinite interval 
at zero density to the one-parameter deformation of the $L$-matrix 
of the Hubbard model.  
By construction, this model has Y(su(2))$\oplus$Y(su(2)) symmetry 
in the infinite chain limit.  Multiparticle eigenstates of the model 
are investigated through this method.




\newpage

\section{Introduction}

As a model of strongly correlated electrons, the Hubbard model has been 
attracting much interest in solid state physics. 
Especially, in one dimension, the model is exactly 
solvable~\cite{Hubbardbook} and 
its thermodynamic properties can be calculated out, which give 
a good testing ground for theories of strongly correlated electron 
systems. From the point of view of the integrability of the one-dimensional 
Hubbard model, there have been many works, including the pioneering 
work of the coordinate Bethe ansatz by Lieb and Wu~\cite{LiWu68}, 
the quantum inverse scattering 
method~\cite{Sha86,Sha88,WaOlAk87,OlWaAk87}, 
its SO(4) invariance~\cite{SO4,EKS,EsKo94}, Y(su(2))$\oplus$Y(su(2)) 
invariance in the infinite chain limit~\cite{UgKo94}, and 
the recent development of the algebraic and analytic Bethe 
ansatz~\cite{RaMa96,YuDe96}.

One of the novel properties of the $R$-matrix $R(\lambda,\mu)$
associated with the one-dimensional Hubbard model is 
that it is thought to be impossible to express it 
as a function of a difference of 
two spectral parameters $\lambda$ and $\mu$. 
This lack of the ``difference property" have prevented us from 
investigating underlying integrable structures of the model.
For example, it is not known whether this $R$-matrix is 
expressible as an intertwiner of a certain algebra.
Since the methods for calculating various correlation functions known so 
far~\cite{KBIbook,JiMiBOOK} requires understanding of such underlying 
structures of the model to some extent, it is necessary to 
deepen our knowledge of the mathematical structures of the Hubbard model 
in order to calculate correlation functions.

The absence of the ``difference property" is not a mere harm for us.
As was noticed by Shiroishi and Wadati, it produces a one-parameter 
integrable extention of the $L$-matrix and the 
Hamiltonian of the Hubbard model~\cite{ShiWa95}.
This extended Hubbard model is interpreted as an electronic model 
with on-site and neighboring-site interactions and 
correlated hopping to the neighboring sites. 
Though the form of the Hamiltonian is complicated and 
thus is difficult to be equipped with physical meaning, 
it can be of some help us with the understanding of the structures 
of the original Hubbard model.

On the other hand, we have recently discovered that the $R$- 
and $L$-matrices of the one-dimensional Hubbard model can be put into
a formulation of quantum inverse scattering method (QISM)
on an infinite interval~\cite{MuGo97,MuGo98}, 
which has been applied to other integrable 
model~\cite{FaTha,CrThWi80}. Through that method, we can 
derive the existence of Yangian symmetry Y(su(2))$\oplus$Y(su(2)) and 
construct $n$-particle states upon zero-density vacuum.
Based on this work, the aim of the present paper is to put the one-parameter 
deformed $L$-matrix, which is 
described in the previous paragraph, into the same 
formulation of 
the QISM on an infinite interval. 
Through this procedure, a new electronic Hamiltonian with 
variable-range correlated hopping arises. It can be embedded in a family 
of an infinite number of commuting operators and thus is 
interpreted as a one-parameter integrable deformation of the
Hubbard chain.
As is the case for the usual Hubbard chain, Yangian invariance of the 
Hamiltonian and construction of multiparticle states can be directly 
established as a byproduct of this method.

This paper is organized as follows.
In Section 2 we shall explain integrability of 
the Hubbard chain of finite length and a one-parameter deformation 
of that model. 
Section 3 is devoted to the application of the QISM on an infinite 
interval to the one-parameter deformation of the $L$-matrix. 
Its resulting new Hamiltonian and commuting conserved operators 
are developed in Section 4. The Yangian  
Y(su(2))$\oplus$Y(su(2)) invariance of the model follows by construction.
In Section 5, we shall construct multiparticle states upon the
zero-density vacuum by use of symmetries and algebras of some 
operators. Section 6 contains concluding remarks and discussions.

\section{Hamiltonian and Monodromy Matrix on the Finite Interval}
\setcounter{equation}{0}

The Hamiltonian for the one-dimensional Hubbard model is
\begin{equation}
\hat{H}=-\sum_{j,\sigma=\uparrow,\downarrow}
(c_{j+1,\sigma}^{\dagger}c_{j,\sigma}+
c_{j,\sigma}^{\dagger}c_{j+1,\sigma})+
U\sum_{j}
\left( n_{j\uparrow}-\frac{1}{2}\right)
\left( n_{j\downarrow}-\frac{1}{2}\right),
\label{eqn:Hamiltonian}
\end{equation}
where $c_{j\sigma}$ and $c_{j\sigma}^{\dagger}$ are respectively the 
fermion annihilation and creation operators which satisfy the 
usual anticommutation relations, and 
$n_{j\sigma}=c_{j\sigma}^{\dagger}c_{j\sigma}$ is the particle number 
operator.
This Hamiltonian has a large symmetry. First, it 
is invariant under the partial particle-hole 
transformation:
\begin{equation}
  \label{eqn:ph}
  c_{j\uparrow}\rightarrow c_{j\uparrow}, \ \ 
  c_{j\downarrow}\rightarrow (-1)^{j}c_{j\downarrow}^{\dagger}, \ \ 
  U\rightarrow -U.
\end{equation}
Second, it is invariant under SO(4) algebra generated by $S^{a}$ and 
$\eta^{a}$ ($a=x,y,z$). $S^{a}$ is the usual spin-SU(2) operator defined by
\begin{equation}
  S^{a}=\frac{1}{2}\sum_{j}\sigma_{\alpha\beta}^{a}
c_{j\alpha}^{\dagger}c_{j\beta},
\end{equation}
and $\eta^{a}$ is obtained by performing the partial particle-hole 
transformation (\ref{eqn:ph}) to $S^{a}$.
If we go to the infinite chain limit, the symmetry is enhanced to 
Y(su(2))$\oplus$Y(su(2)) Yangian~\cite{UgKo94}.

The integrability of the fermionic Hubbard model (\ref{eqn:Hamiltonian})
is based on a 
local exchange relation~\cite{OlWaAk87}
\begin{equation}
{\check{\cal R}}(\lambda,\mu)[ {\cal L}_{j}(\lambda)
\otimes_{{\rm s}} {\cal L}_{j}(\mu) ]
=[ {\cal L}_{j}(\mu)\otimes_{{\rm s}} {\cal L}_{j}(\lambda) ]
{\check {\cal R}}(\lambda,\mu).
\label{eqn:RLL}
\end{equation}
where
$\otimes_{{\rm s}}$ denotes the Grassman tensor product
\begin{equation}
[A\otimes_{{\rm s}}B]_{\alpha\gamma,\beta\delta}
=(-1)^{[P(\alpha)+P(\beta)]P(\gamma)}A_{\alpha\beta}B_{\gamma\delta},
\end{equation}
with the grading $P(1)=P(4)=0$, $P(2)=P(3)=1$.
The expressions for the matrices ${\check {\cal R}}$ and ${\cal L}$ are 
presented in \ref{appendix:RL}.\footnote{The matrix ${\cal R}$ 
in Ref.~\cite{OlWaAk87,MuGo97,MuGo98} is written as ${\check {\cal R}}$ 
in this paper, following the standard notation. It should not be 
confused with the matrix written in the standard notation as 
${\cal R}={\cal P}{\check {\cal R}}$.} 
In these expressions, we use a function $h(\lambda)$ defined by
\begin{equation}
\frac{\sinh 2h(\lambda)}{\sin 2\lambda}=\frac{U}{4}.
\label{eqn:U4}
\end{equation}
The Hamiltonian (\ref{eqn:Hamiltonian}) 
is reproduced by a logarithmic derivative of 
the transfer matrix $\tau_{mn}(\lambda)$;
\begin{equation}
\hat{H}=\frac{d}{d\lambda}\ln\tau_{mn}(\lambda)|_{\lambda=0}, \ \ \ 
\tau_{mn}(\lambda)={\rm str}({\cal T}_{mn}(\lambda))={\rm tr}
((\sigma^{z}\otimes\sigma^{z}){\cal T}_{mn}(\lambda)),
\end{equation}
where the monodromy matrix ${\cal T}$ is given by
\begin{equation}
{\cal T}_{mn}(\lambda)={\cal L}_{m-1}(\lambda){\cal L}_{m-2}
(\lambda)\cdots {\cal L}_{n}(\lambda) \ \  (m>n).
\label{eqn:finiteT}
\end{equation}

There is known to be an integrable spin chain equivalent to 
the Hubbard model~\cite{Sha86,Sha88}.
If we apply the Jordan-Wigner transformation 
\begin{equation}
c_{j\uparrow}=\sigma_{n}^{z}\cdots\sigma_{j-1}^{z}\sigma_{j}^{-}, 
\ \ \ 
c_{j\downarrow}=(\sigma_{n}^{z}\cdots\sigma_{m-2}^{z}\sigma_{m-1}^{z})
\tau_{n}^{z}\cdots \tau_{j-1}^{z}\tau_{j}^{-},
\end{equation}
where $\sigma$ and $\tau$ are the Pauli matrices, and 
$
\sigma_{j}^{\pm}=(\sigma_{j}^{x}\pm{\rm i}\sigma_{j}^{y})/2,
\ \
\tau_{j}^{\pm}=(\tau_{j}^{x}\pm{\rm i}\tau_{j}^{y})/2, $
we get an equivalent spin model 
\begin{equation}
\hat{H}=
\sum_{j}(\sigma_{j+1}^{+}\sigma_{j}^{-}+
\sigma_{j}^{+}\sigma_{j+1}^{-})
+\sum_{j}(\tau_{j+1}^{+}\tau_{j}^{-}+
\tau_{j}^{+}\tau_{j+1}^{-})
+\frac{U}{4}\sum_{j}\sigma_{j}^{z}\tau_{j}^{z}.
\label{eqn:spinHamiltonian}
\end{equation}
Its integrability is supported by the spin-chain counterpart of the
exchange relation~\cite{Sha88}
\begin{equation}
{\check R}(\lambda,\mu)[  L_{j}(\lambda)\otimes L_{j}(\mu) ]
=[ L_{j}(\mu)\otimes L_{j}(\lambda) ]{\check R}(\lambda,\mu).
\label{eqn:spinRLLcheck}
\end{equation}
By using $R_{12}(\lambda,\mu)=P_{12}{\check R}_{12}(\lambda,\mu)$, 
where $P_{12}$ is the transposition $P(x\otimes y)=y\otimes x$, 
it is also written as 
\begin{equation}
R_{12}(\lambda,\mu)L_{1j}(\lambda)L_{2j}(\mu) 
=L_{2j}(\mu)L_{1j}(\lambda)R_{12}(\lambda,\mu),
\label{eqn:spinRLL}
\end{equation}
As Olmedilla et al.~\cite{OlWaAk87} found, 
the exchange relation of the fermionic model, (\ref{eqn:RLL}), 
and that of the spin chain, (\ref{eqn:spinRLLcheck}),
can be transformed into 
each other. 

One of the peculiarity on the integrability of the 
Hubbard model known for years is that the ${\check {\cal R}}$-matrix,
or equivalently the ${\check R}$-matrix,
is believed to lack the difference property, 
i.e. it is not a function of $\lambda-\mu$, nor
can it be expressed as $f(\lambda)-f(\mu)$ with some function 
$f$. This has been an obstruction for further investigations of 
the underlying mathematical structures of the Hubbard model.
But on the other hand, the lack of the difference property allows 
us to consider a one-parameter integrable deformation of the Hubbard 
model, as noticed by Shiroishi and Wadati~\cite{ShiWa95}. 
This works as follows.
Since the $R$-matrix is shown to satisfy the Yang-Baxter equation of the form
\begin{equation}
R_{12}(\lambda,\mu)R_{13}(\lambda,\nu)R_{23}(\mu,\nu)
=R_{23}(\mu,\nu)R_{13}(\lambda,\nu)R_{12}(\lambda,\mu),
\end{equation}
a new $L$-matrix defined by 
\begin{equation}
L_{1}(\lambda)_{\nu}=R_{13}(\lambda,\nu)
\end{equation}
satisfies an exchange relation
\begin{equation}
R_{12}(\lambda,\mu)L_{1}(\lambda)_{\nu} L_{2}(\mu)_{\nu}
=L_{2}(\mu)_{\nu}L_{1}(\lambda)_{\nu}R_{12}(\lambda,\mu).
\label{eqn:spinRLLnu}
\end{equation}
By using ${\check R}_{12}(\lambda,\mu)=P_{12}R_{12}(\lambda,\mu)$, 
it can be alternatively written as  
\begin{equation}
{\check R}(\lambda,\mu)[L(\lambda)_{\nu} \otimes L(\mu)_{\nu}]
=[L(\mu)_{\nu}\otimes L(\lambda)_{\nu}]{\check R}(\lambda,\mu).
\label{eqn:checkRLLnu}
\end{equation}
Considering that $L(\lambda)_{\nu=0}\propto L(\lambda)$, which can be 
checked by a direct calculation, 
we can say that $L(\lambda)_{\nu}$ is a one-parameter 
deformation of the $L$-matrix $L(\lambda)$ 
of the original Hubbard model. 
This new $L$-matrix can be used to produce a new 
Hamiltonian~\cite{ShiWa95}.
By using the monodromy matrix given by 
\begin{equation}
T_{mn}(\lambda)_{\nu}=L_{m-1}(\lambda)_{\nu}
L_{m-2}(\lambda)_{\nu}\cdots 
L_{n}(\lambda)_{\nu} \ \ (m>n),
\end{equation}
the new Hamiltonian is given by 
\begin{eqnarray}
&&\hat{H}_{\nu}=\left.\frac{d}{d\lambda}\ln
{\rm tr}(T_{mn}(\lambda)_{\nu})
\right|_{\lambda=\nu} \nonumber \\
&&\ \ =
-\sum_{j,\sigma=\uparrow,\downarrow}
(c_{j+1,\sigma}^{\dagger}c_{j,\sigma}+
c_{j,\sigma}^{\dagger}c_{j+1,\sigma})\nonumber \\
&& \ \ \ +
\frac{U}{4\cosh 2h(\nu)} \sum_{j}
\left( 
2n_{j\uparrow}\cos^{2}\nu
-2n_{j+1\uparrow}\sin^{2}\nu
+\sin2\nu(c^{\dagger}_{j\uparrow}c_{j+1\uparrow}-
c^{\dagger}_{j+1\uparrow}c_{j\uparrow})-\cos 2\nu
\right) \nonumber \\
&& \ \ \ \ 
\left( 
2n_{j\downarrow}\cos^{2}\nu
-2n_{j+1\downarrow}\sin^{2}\nu
+\sin2\nu(c^{\dagger}_{j\downarrow}c_{j+1\downarrow}-
c^{\dagger}_{j+1\downarrow}c_{j\downarrow})-\cos 2\nu
\right).
\label{eqn:Hamiltoniannu}
\end{eqnarray}
We have performed the Jordan-Wigner transformation to 
get a fermionic Hamiltonian. The reason for choosing the special 
value $\lambda=\nu$ is to 
obtain a local Hamiltonian.
This Hamiltonian is Hermitian if $\nu$ is pure imaginary, 
and in that case it represents a model with 
on-site and neighoboring-site 
interaction and correlated hopping to the neighboring sites.
The exchange relation (\ref{eqn:checkRLLnu}) results in 
\begin{equation}
[\ln{\rm tr}T_{mn}(\lambda)_{\nu},
\ln{\rm tr}T_{mn}(\mu)_{\nu}]=0
\end{equation}
Therefore, the Hamiltonian is embedded in a family of infinite number 
of commuting operators, and so the model (\ref{eqn:Hamiltoniannu}) 
is integrable. The model (\ref{eqn:Hamiltoniannu}) includes
the Hubbard model (\ref{eqn:Hamiltonian}) as the $\nu=0$ case. 

So far we have discussed on a spin-chain version of the 
exchange relation (\ref{eqn:checkRLLnu}).
We can employ the same procedure used in Ref.~\cite{OlWaAk87} to 
fermionize the exchange relation of the spin chain 
(\ref{eqn:checkRLLnu}).
The Jordan-Wigner transformation relates the spin operators
and the fermion operators as 
\begin{equation}
\left(
\begin{array}{c}
\sigma_{j}^{+} \\ \sigma_{j}^{-}
\end{array}
\right)
=V_{j\uparrow}^{2}
\left(
\begin{array}{c}
c_{j\uparrow}^{\dagger} \\ c_{j\uparrow}
\end{array}
\right), \ \ \ 
\left(
\begin{array}{c}
\tau_{j}^{+} \\ \tau_{j}^{-}
\end{array}
\right)
=V_{j\downarrow}^{2}
\left(
\begin{array}{c}
c_{j\downarrow}^{\dagger} \\ c_{j\downarrow}
\end{array}
\right).
\end{equation}
Here the matrices $V_{j\uparrow}$ and $V_{j\downarrow}$ are defined 
by~\cite{OlWaAk87}
\begin{eqnarray}
&&V_{j\sigma}=\left(
\begin{array}{cc}
\exp\left( {\rm i}\frac{\pi}{2}
\Omega_{j\sigma}\right) &0 \\0 & 
\exp\left( -{\rm i}\frac{\pi}{2}
\Omega_{j\sigma}\right)
\end{array}
\right) .\\
&&
\Omega_{j\uparrow}=
\sum_{k=n}^{j-1}(n_{k\uparrow}-1), \  \ \ 
\Omega_{j\downarrow}=
\sum_{k=n}^{m-1}(n_{k\uparrow}-1)+
\sum_{k=n}^{j-1}(n_{k\downarrow}-1).
\end{eqnarray}
Then we obtain the fermionic exchange relation;
\begin{equation}
{\check {\cal R}}(\lambda,\mu)[ {\cal L}_{j}(\lambda)_{\nu}
\otimes_{{\rm s}} {\cal L}_{j}(\mu)_{\nu} ]
=[ {\cal L}_{j}(\mu)_{\nu}\otimes_{{\rm s}} {\cal L}_{j}(\lambda)_{\nu} ]
{\check {\cal R}}(\lambda,\mu),
\label{eqn:RLLnu}
\end{equation}
where the fermionic matrix ${\cal L}_{j}(\lambda)$ can be calculated
from the spin-chain matrix $L_{j}(\lambda)$ as 
\begin{equation}
{\cal L}_{j}(\lambda)_{\nu}=(V_{j+1,\uparrow}\otimes V_{j+1,\downarrow})
L_{j}(\lambda)_{\nu}(V_{j\uparrow}^{-1}\otimes V_{j\downarrow}^{-1}),
\end{equation}
and the corresponding ${\cal R}$-matrix is
\begin{equation}
{\check {\cal R}}(\lambda,\mu)=W{\check R}(\lambda,\mu)W^{-1},
\end{equation}
with
$W=\sigma^{z}\otimes
{\rm diag}(
1,-{\rm i},-{\rm i},1)
\otimes I_{2}$. Here $I_{n}$ denotes the $n\times n$ unit matrix. 
The explicit form of ${\cal L}_{j}(\lambda)_{\nu}$ is presented in 
\ref{appendix:RL}.
The exchange relation (\ref{eqn:checkRLLnu}) and (\ref{eqn:RLLnu}) 
can be considered as a 
one-parameter deformation of (\ref{eqn:spinRLLcheck}) and (\ref{eqn:RLL}),
respectively. Note that ${\check R}(\lambda,\mu)$ or 
${\check {\cal R}}(\lambda,\mu)$ is unchanged by this one-parameter 
deformation.

\section{Passage to the Infinite Interval}
\label{sec:passage}
\setcounter{equation}{0}

Next we pass to the infinite interval limit using the new exchange 
relation (\ref{eqn:RLLnu}). 
The method is identical with the one in our previous works 
on the original Hubbard model~\cite{MuGo97,MuGo98}.
Let 
\begin{eqnarray}
{\cal T}_{mn}(\lambda)_{\nu}&=&
{\cal L}_{m-1}(\lambda)_{\nu}
{\cal L}_{m-2}(\lambda)_{\nu}\cdots
{\cal L}_{n}(\lambda)_{\nu}, \\
{\cal T}_{mn}^{(2)}
(\lambda,\mu)_{\nu}&=&{\cal T}_{mn}(\lambda)_{\nu}\otimes_{{\rm s}}
{\cal T}_{mn}(\mu)_{\nu}.
\end{eqnarray}
${\cal T}_{mn}(\lambda)_{\nu}$ is a monodromy matrix on the finite interval.
To consider the infinite-chain limit of the monodromy matrix,
we should split off the asymptotics of its vacuum expectation value 
for $m,-n\rightarrow\infty$. 
Hence, this procedure is restricted to uncorrelated vacua, 
with which one can calculate vacuum 
expectation value of the monodromy matrix.
Among four uncorrelated vacua, in which the electron density of  
each spin is either zero or unity, 
we take the zero-density vacuum $|0\rangle$ 
to renormalize the monodromy matrix.
We use the following two matrices
\begin{equation}
V(\lambda)_{\nu}=
\langle 0|{\cal L}_{j}(\lambda)_{\nu}|0\rangle, \ \ \ 
V^{(2)}(\lambda,\mu)_{\nu}=\langle 0|{\cal L}_{j}(\lambda)_{\nu}
\otimes_{{\rm s}}
{\cal L}_{j}(\mu)_{\nu}|0\rangle . 
\end{equation}
in order to normalize ${\cal T}_{mn}(\lambda)$ and 
${\cal T}_{mn}^{(2)}(\lambda,\mu)$, respectively;
\begin{eqnarray}
\tilde{{\cal T}}(\lambda)_{\nu}&=&\lim_{m,-n\rightarrow\infty}
V(\lambda)_{\nu}^{-m}
{\cal T}_{mn}(\lambda)_{\nu}V(\lambda)_{\nu}^{n}, \\
\tilde{{\cal T}}^{(2)}(\lambda,\mu)_{\nu}&=&\lim_{m,-n\rightarrow\infty}
V^{(2)}(\lambda,\mu)_{\nu}^{-m}
{\cal T}_{mn}^{(2)}(\lambda,\mu)_{\nu}V^{(2)}(\lambda,\mu)_{\nu}^{n}.
\end{eqnarray}
These limits converge in the weak sense, 
though the matrices ${\cal T}_{mn}(\lambda)_{\nu}$ and 
${\cal T}_{mn}^{(2)}(\lambda,\mu)_{\nu}$ do not have a definite 
limit when $m,-n\rightarrow\infty$.
We shall call $\tilde{{\cal T}}(\lambda)_{\nu}$ a monodromy matrix on 
the infinite interval. It allows an alternative definition:
\begin{equation}
\tilde{{\cal T}}(\lambda)_{\nu}
=I_{4}+\sum_{j}(\tilde{{\cal L}}_{j}(\lambda)_{\nu}-I_{4}) 
 +\sum_{j>i}(\tilde{{\cal L}}_{j}(\lambda)_{\nu}-I_{4}) 
(\tilde{{\cal L}}_{i}(\lambda)_{\nu}-I_{4})+\cdots ,
\label{eqn:tildeTsum}
\end{equation}
where $\tilde{{\cal L}}_{j}(\lambda)_{\nu}=V(\lambda)_{\nu}^{-j-1}
{\cal L}_{j}(\lambda)_{\nu}V(\lambda)_{\nu}^{j}$.

For practical calculations, one should 
be careful that $V^{(2)}(\lambda,\mu)_{\nu}$ is not equal to the 
tensor product $V(\lambda)_{\nu}\otimes_{{\rm s}}V(\mu)_{\nu}$. 
There appear additional off-diagonal elements due to normal ordering of
operators. 
Direct calculations lead us to the resulting forms for 
$V(\lambda)_{\nu}$ and  $V^{(2)}(\lambda,\mu)_{\nu}$;
\[
V(\lambda)_{\nu}=
{\rm diag} 
(-\rho_{8}(\lambda,\nu),\rho_{9}(\lambda,\nu),
\rho_{9}(\lambda,\nu),
-\rho_{1}(\lambda,\nu)).
\label{eqn:defV}
\]
As for the matrix $V^{(2)}(\lambda,\mu)_{\nu}$, 
its diagonal consists of the elements of 
$V(\lambda)_{\nu}\otimes_{{\rm s}}V(\mu)_{\nu}$, and its 
non-vanishing off-diagonal elements are 
\begin{eqnarray*}
&&V^{(2)}(\lambda,\mu)_{\nu}^{12,21}=V^{(2)}(\lambda,\mu)_{\nu}^{13,31}=
-{\rm i}\rho_{6}(\lambda,\nu)\rho_{6}(\mu,\nu), \\
&&V^{(2)}(\lambda,\mu)_{\nu}^{14,23}=-V^{(2)}(\lambda,\mu)_{\nu}^{14,32}=
-{\rm i}\rho_{6}(\lambda,\nu)\rho_{2}(\mu,\nu), \\
&&V^{(2)}(\lambda,\mu)_{\nu}^{24,42}=V^{(2)}(\lambda,\mu)_{\nu}^{34,43}=
{\rm i}\rho_{2}(\lambda,\nu)\rho_{2}(\mu,\nu), \\
&&V^{(2)}(\lambda,\mu)_{\nu}^{23,41}=-V^{(2)}(\lambda,\mu)_{\nu}^{32,41}=
-{\rm i}\rho_{2}(\lambda,\nu)\rho_{6}(\mu,\nu), \\
&&V^{(2)}(\lambda,\mu)_{\nu}^{14,41}=-\rho_{3}(\lambda,\nu)\rho_{3}(\mu,\nu).
\end{eqnarray*}
Note that $V^{(2)}(\lambda,\mu)_{\nu}$ is upper triangular. Since the 
diagonals of $V^{(2)}(\lambda,\mu)_{\nu}$ and $V(\lambda)\otimes_{{\rm s}}
V(\mu)_{\nu}$ are identical, $V^{(2)}(\lambda,\mu)_{\nu}$ can be diagonalized 
by an upper 
triangular matrix $U(\lambda,\mu)$ whose diagonal elements are all 
unity;
\begin{equation}
  V^{(2)}(\lambda,\mu)_{\nu}=U(\lambda,\mu)_{\nu}(
V(\lambda)_{\nu}\otimes_{{\rm s}}
V(\mu)_{\nu})U(\lambda,\mu)_{\nu}^{-1}.
\label{eqn:VUVVU}
\end{equation}
Direct calculation leads us to 
a remarkable and surprising fact; $U(\lambda,\mu)_{\nu}$ is 
independent of $\nu$.
It is equal to $U(\lambda,\mu)_{\nu=0}=U(\lambda,\mu)$, which has 
appeared 
in the analysis of the usual Hubbard chain~\cite{MuGo97}, so 
we will hereafter suppress the subscript $\nu$ in $U(\lambda,\mu)_{\nu}$.
Its matrix elements are
\begin{eqnarray*}
&& U(\lambda,\mu)_{12,21}=
U(\lambda,\mu)_{13,31}=-{\rm i}\rho_{2}/\rho_{10}, \ \ \
U(\lambda,\mu)_{14,23}=
-U(\lambda,\mu)_{14,32}={\rm i}\rho_{6}/\rho_{8}, \\
&&U(\lambda,\mu)_{24,42}=
U(\lambda,\mu)_{34,43}={\rm i}\rho_{2}/\rho_{9}, \ \ \ 
U(\lambda,\mu)_{23,41}=
-U(\lambda,\mu)_{32,41}={\rm i}\rho_{6}/\rho_{7},\\
&&U(\lambda,\mu)_{14,41}=-\rho_{5}/\rho_{7},
\end{eqnarray*}
where $\rho_{i}=\rho_{i}(\lambda,\mu)$.

Taking the vacuum expectation value of the local exchange relation
(\ref{eqn:RLL}) yields
\begin{equation}
{\check {\cal R}}(\lambda,\mu)V^{(2)}(\lambda,\mu)_{\nu}=
V^{(2)}(\mu,\lambda)_{\nu}
{\check {\cal R}}(\lambda,\mu),
\end{equation}
and we conclude that 
\begin{equation}
{\check {\cal R}}(\lambda,\mu)\tilde{{\cal T}}^{(2)}(\lambda,\mu)_{\nu}=
\tilde{{\cal T}}^{(2)}(\mu,\lambda)_{\nu}
{\check {\cal R}}(\lambda,\mu).
\label{eqn:exchangeT2}
\end{equation}

Finally, collecting (\ref{eqn:exchangeT2}) and other equations together, 
we arrive at the exchange relation for the monodromy matrix 
$\tilde{{\cal T}}(\lambda)_{\nu}$ on the infinite interval, 
\begin{equation}
\tilde{{\cal R}}^{(+)}(\lambda,\mu)\left[
\tilde{\cal T}(\lambda)_{\nu}\otimes_{{\rm s}}\tilde{\cal T}(\mu)_{\nu}\right]
=\left[\tilde{\cal T}(\mu)_{\nu}
\otimes_{{\rm s}}\tilde{\cal T}(\lambda)_{\nu}\right]
\tilde{{\cal R}}^{(-)}(\lambda,\mu),
\label{eqn:newRTT}
\end{equation}
where
\begin{eqnarray}
&&\tilde{\cal R}^{(\pm)}(\lambda,\mu)_{\nu}=
U_{\pm}(\mu,\lambda)_{\nu}^{-1}{\check {\cal R}}(\lambda,\mu)
U_{\pm}(\lambda,\mu)_{\nu},
\label{eqn:tildeRpm} \\
&& U_{\pm}(\lambda,\mu)_{\nu}=\lim_{m\rightarrow\pm\infty}
V^{(2)}(\lambda,\mu)_{\nu}^{-m}
[ V(\lambda)_{\nu}^{m}\otimes_{s}V(\mu)_{\nu}^{m}].
\label{eqn:Upm}
\end{eqnarray}
Since the calculation of the matrices $U_{\pm}(\lambda,\mu)_{\nu}$
and $\tilde{\cal R}^{(\pm)}(\lambda,\mu)_{\nu}$ is rather technical, 
we do not reproduce it here. Its details are presented in Ref.~\cite{MuGo98}
in the case of $\nu=0$ (Hubbard model).
The only point we should note here is that apart from some 
singular points (e.g. $\lambda=\mu$), we can say that 
$U(\lambda,\mu)=U_{\pm}(\lambda,\mu)_{\nu}$ and 
\begin{eqnarray}
\lefteqn{\tilde{\cal R}(\lambda,\mu) =
\tilde{\cal R}^{(\pm)}(\lambda,\mu)_{\nu}
= }\nonumber \\[1ex]
& & \hspace{-22pt} \left( 
{\arraycolsep 2pt
\begin{array}{cccccccccccccccc}
\rho_{1} & 0 & 0 & 0 & 
0 & 0 & 0 & 0 & 0 & 0 & 0 & 0 & 0 & 0 & 0 & 0 \\
0 & 0 & 0 & 0 & 
\frac{\rho_{1}\rho_{4}}{{\rm i}\rho_{10}} 
& 0 & 0 & 0 & 0 & 0 & 0 & 0 
& 0 & 0 & 0 & 0 \\
0 & 0 & 0 & 0 & 
0 & 0 & 0 & 0 &
\frac{\rho_{1}\rho_{4}}{{\rm i}\rho_{10}} 
& 0 & 0 & 0 & 0 & 0 & 0 & 0 \\
0 & 0 & 0 & 0 & 
0 & 0 & 0 & 0 &
0 & 0 & 0 & 0 &
\frac{-\rho_{1}\rho_{4}}{\rho_{5}-\rho_{4}} & 0 & 0 & 0 \\
0 & 
 -{\rm i}\rho_{10} &0 & 0 & 0 &  0 & 0 & 0 &
0 & 0 & 0 & 0 &
0 & 0 & 0 & 0 \\
0 & 0 & 0 & 0 & 
0 &  \rho_{4} & 0 & 0 &
0 & 0 & 0 & 0 &
0 & 0 & 0 & 0 \\
0 & 0 & 0 & 0 & 
0 & 0 & \frac{\rho_{3}\rho_{4}-\rho_{2}^{2}}{\rho_{3}-\rho_{1}}
& 0 & 0 & 
\frac{\rho_{9}\rho_{10}}{\rho_{3}-\rho_{1}} & 0 & 0 &
0 & 0 & 0 & 0 \\
0 & 0 & 0 & 0 & 
0 & 0 & 0 & 0 &
0 & 0 & 0 & 0 &
0 & 
\frac{{\rm i}\rho_{1}\rho_{4}}{\rho_{9}}
 & 0 & 0 \\
0 & 0 &  -{\rm i}\rho_{10} & 0 & 
0 & 0 & 0 & 0 &
0 & 0 & 0 & 0 &
0 & 0 & 0 & 0 \\
0 & 0 & 0 & 0 & 
0 & 0 &\frac{\rho_{9}\rho_{10}}{\rho_{3}-\rho_{1}}
& 0 & 0 &  \frac{\rho_{3}\rho_{4}-\rho_{2}^{2}}{\rho_{3}-\rho_{1}}
 & 0 & 0 & 
0 & 0 & 0 & 0 \\
0 & 0 & 0 & 0 & 
0 & 0 & 0 & 0 &
0 & 0 &  \rho_{4} & 0 &
0 & 0 & 0 & 0 \\
0 & 0 & 0 & 0 & 
0 & 0 & 0 & 0 &
0 & 0 & 0 & 0 &
0 & 0 & \frac{{\rm i}\rho_{1}\rho_{4}}{\rho_{9}}
& 0 \\
0 & 0 & 0 &  \rho_{1}-\rho_{3} & 
0 & 0 & 0 & 0 &
0 & 0 & 0 & 0 &
0 & 0 & 0 & 0 \\
0 & 0 & 0 & 0 & 
0 & 0 & 0 &  {\rm i} \rho_{9} &
0 & 0 & 0 & 0 &
0 & 0 & 0 & 0 \\
0 & 0 & 0 & 0 & 
0 & 0 & 0 & 0 &
0 & 0 & 0 &  {\rm i} \rho_{9} &
0 & 0 & 0 & 0 \\
0 & 0 & 0 & 0 & 
0 & 0 & 0 & 0 &
0 & 0 & 0 & 0 &
0 & 0 & 0 &  \rho_{1} 
\end{array}
}
\right)
\raisebox{-22ex}{,} \makebox[2em]{}
\label{eqn:deftildeR}
\end{eqnarray}
where $\rho_{i}=\rho_{i}(\lambda,\mu)$.
Let us write the elements of $\tilde{{\cal T}}(\lambda)$ as 
\begin{equation}
\tilde{{\cal T}}(\lambda)_{\nu}=
\left(
\begin{array}{cccc}
D_{11}(\lambda)_{\nu} & C_{11}(\lambda)_{\nu} 
&C_{12}(\lambda)_{\nu} & D_{12}(\lambda)_{\nu} \\
B_{11}(\lambda)_{\nu} & A_{11}(\lambda)_{\nu} 
&A_{12}(\lambda)_{\nu} & B_{12}(\lambda)_{\nu} \\
B_{21}(\lambda)_{\nu} & A_{21}(\lambda)_{\nu} 
&A_{22}(\lambda)_{\nu} & B_{22}(\lambda)_{\nu} \\
D_{21}(\lambda)_{\nu} & C_{21}(\lambda)_{\nu} 
&C_{22}(\lambda)_{\nu} & D_{22}(\lambda)_{\nu} 
\end{array} 
\right).
\label{eqn:matrixT}
\end{equation}
Since the $\tilde{{\cal R}}$-matrix is independent of the 
value of $\nu$, the commutation rules between the elements 
of $\tilde{{\cal T}}(\lambda)$, which are obtained from 
the exchange relation (\ref{eqn:newRTT}), are completely the 
same as the ones in the $\nu=0$ case, i.e. the usual Hubbard model. 
The complete list of the commutation rules are 
found in Appendix B
 of Ref.~\cite{MuGo98}, and it is the same in the present case.

\section{Yangian Symmetry and Commuting Operators}
\setcounter{equation}{0}

If we follow the notion of the quantum inverse scattering method, 
the remaining task is to investigate the meaning of each matrix element
of $\tilde{{\cal T}}(\lambda)_{\nu}$. 
As is also the case for the Hubbard model~\cite{MuGo98},
the commutation relations between the elements of the submatrix 
$A(\lambda)_{\nu}$ decouple from the rest of the algebra.
\begin{equation} 
     r(\lambda, \mu) \, (A (\lambda)_{\nu} \otimes A (\mu)_{\nu}) =
        (A(\mu)_{\nu} \otimes A (\lambda)_{\nu}) \, r(\lambda, \mu) ,
\label{eqn:commutA}
\end{equation}
where
\begin{equation}
     r(\lambda,\mu) = \frac{\rho_{3}\rho_{4}-\rho_{2}^{2}+\rho_{9}\rho_{10}
{\cal P}}{\rho_{4}
(\rho_{3}-\rho_{1})}  \ \  \ \ \ (\rho_{j}=\rho_{j}(\lambda,\mu)),
\label{eqn:matR}
\end{equation}
and ${\cal P}$ is a $4\times 4$ permutation matrix 
(${\cal P}x\otimes y=y\otimes x$).
As is remarked previously, (\ref{eqn:commutA}) is 
identical with the one in the Hubbard model ($\nu=0$) and 
we can follow the same argument as in the previous work~\cite{MuGo98}. 
By the reparameterization
\begin{equation}
v(\lambda)=-2{\rm i} \, \cot 2 \lambda\cosh 2h(\lambda),
\label{eqn:defv}
\end{equation}
the $R$-matrix $r(\lambda,\mu)$ turns into the rational $R$-matrix 
of the XXX spin chain,
\begin{equation}
r(\lambda,\mu)=\frac{{\rm i}U+(v(\lambda)-v(\mu)){\cal P}}{
{\rm i}U+v(\lambda)-v(\mu)}.
\end{equation}

Let us expand $A(\lambda)_{\nu}$ in terms of $v(\lambda)^{-1}$, 
\begin{equation}
A(\lambda)_{\nu}=I_{2}+{\rm i}U \sum_{n=0}^{\infty}
\frac{1}{v(\lambda)^{n+1}}
\left( \sum_{a=1}^{3}
Q_{n}^{a}(\nu)\tilde{\sigma}^{a}+Q_{n}^{0}(\nu)I_{2}
\right),
\label{eqn:expandA}
\end{equation}
where $\tilde{\sigma}^{x}=-\sigma^{y}$, $\tilde{\sigma}^{y}=\sigma^{x}$, 
and $\tilde{\sigma}^{z}=\sigma^{z}$.
This expansion can be achieved by considering 
the limit $v(\lambda)\rightarrow\infty$ 
as $\Im (\lambda) \rightarrow \infty$ 
and by choosing the proper branch of 
solution of eq.~(\ref{eqn:U4}), which determines $h$ as a function of
$\lambda$. 
(\ref{eqn:U4}) implies that
\begin{equation}
{\rm e}^{-2h(\lambda)}=-\frac{U}{4}\sin 2\lambda \pm 
\sqrt{1+\left( \frac{U}{4}\sin 2\lambda \right)^{2} }.
\label{eqn:e2h}
\end{equation}
To achieve convergence of the matrix elements $A(\lambda)_{\nu}$,
we have to choose the lower sign here.
Then it follows from general considerations~\cite{BGHP,Hal94,MuWa96b}
that the first six operators $Q_{0}^{a}(\nu),Q_{1}^{a}(\nu)$ generate a 
representation of the Y(su(2)) Yangian quantum group.

There is an alternative description of the Yangian 
Y(su(2))~\cite{Dri86} described below.
The Yangian Y(su(2)) is a Hopf algebra spanned by six 
generators $Q_{n}^{a} \ (n=0,1, \ a=x,y,z)$, satisfying the 
following relations, 
\begin{eqnarray}
& & \left[ Q_{0}^{a}, Q_{0}^{b} \right]  =  f^{abc}Q_{0}^{c}, 
\label{eqn:Y1} \\
& & \left[ Q_{0}^{a}, Q_{1}^{b} \right]  =  f^{abc}Q_{1}^{c}, 
\label{eqn:Y2} \\
& & \left[ \left[ Q_{1}^{a},Q_{1}^{b} \right], 
\left[ Q_{0}^{c},Q_{1}^{d} \right] \right]+
\left[ \left[ Q_{1}^{c},Q_{1}^{d} \right], 
\left[ Q_{0}^{a},Q_{1}^{b} \right] \right] \nonumber \\
& & \makebox[2em]{}
=\kappa^{2}(A^{abkefg}f^{cdk}+A^{cdkefg}f^{abk})
\{Q_{0}^{e}, Q_{0}^{f}, Q_{1}^{g} \}.
\label{eqn:Y3}
\end{eqnarray}
Here $\kappa$ is a nonzero constant, 
$f^{abc}=i\varepsilon^{abc}$ is the antisymmetric tensor of 
structure constants of su(2), and 
$A^{abcdef}=f^{adk}f^{bel}f^{cfm}f^{klm}$.
The bracket $\{\; \; \}$ in (\ref{eqn:Y3}) denotes the 
symmetrized product 
\begin{equation}
\{ x_{1}, x_{2}, x_{3}\} =\frac{1}{3!} \sum_{\sigma\in S_{3}}
x_{\sigma 1}x_{\sigma 2}x_{\sigma 3}.
\end{equation}
The Hopf algebra structure of Y(su(2)) is described in 
ref.~\cite{Dri86} and its 
representation theory, which will be used later, is 
developed in Ref.~\cite{ChaPre,ChaPreBOOK}.

We can use (\ref{eqn:expandA}) and 
(\ref{eqn:tildeTsum}) to get 
the representation of Yangian generators;
\begin{eqnarray}
Q_{0}^{a}(\nu)&=& \frac{1}{2}\sum_{j}\sigma_{\alpha\beta}^{a}
c^{\dagger}_{j,\alpha}c_{j,\beta}, 
\label{eqn:Q0}
\\
Q_{1}^{a}(\nu)&=& \frac{{\rm i}}{2\sin\nu\cos\nu}
\left[
\sum_{i>j}(\tan\nu )^{i-j}{\rm e}^{2h(\nu)(2-n_{i}-n_{j})}
\sigma_{\alpha\beta}^{a}c_{i\alpha}^{\dagger}c_{j\beta}\right. 
\nonumber \\
&&\makebox[3em]{}
+\left.\sum_{i<j}(-\cot\nu )^{i-j}{\rm e}^{-2h(\nu)(2-n_{i}-n_{j})}
\sigma_{\alpha\beta}^{a}c_{i\alpha}^{\dagger}c_{j\beta}
\right] \nonumber\\
&&\makebox[1em]{}
-\frac{{\rm i}U}{4}\sum_{i,j}
{\rm sgn}(j-i)\sigma_{\alpha\beta}^{a}
c_{i,\alpha}^{\dagger}c_{j,\gamma}^{\dagger}
c_{i,\gamma}c_{j,\beta}+{\rm i}U{\rm cot}2h(\nu) \ \sin^{2}\nu \ Q_{0}^{a}
(\nu).
\label{eqn:Q1}
\end{eqnarray}
In this case the constant $\kappa$ in (\ref{eqn:Y3}) is equal to 
${\rm i}U$.
Note that $Q_{0}^{a}(\nu)=S^{a}$ is just the operator of the 
$a$-component of the total spin. 
The representation of the Yangian algebra in the usual Hubbard 
model~\cite{UgKo94,MuGo97} is 
a special case of $\nu=0$ in (\ref{eqn:Q0}) and (\ref{eqn:Q1}).

Since the quantum determinant
\begin{equation}
{\rm Det}_{q}A(\lambda)_{\nu}=
A_{11}(\lambda)_{\nu}A_{22}(\check{\lambda})_{\nu}-
A_{12}(\lambda)_{\nu}A_{21}(\check{\lambda})_{\nu},
\end{equation}
where $v(\check{\lambda})=v(\lambda)-{\rm i}U$,
is in the center of the Yangian 
\begin{equation}
[{\rm Det}_{q}A(\lambda)_{\nu}, A(\mu)_{\nu}]=0, 
\label{eqn:DetAA}
\end{equation}
it provides a generating
function of mutually commuting operators,
\begin{equation}
[{\rm Det}_{q}A(\lambda)_{\nu}, {\rm Det}_{q}A(\mu)_{\nu}]=0.
\label{eqn:DetADetA}
\end{equation}
The asymptotic expansion in
terms of $v(\lambda)^{-1}$, 
\begin{equation}
{\rm Det}_{q}A(\lambda)_{\nu}= 1+{\rm i}U\sum_{n=0}^{\infty}
\frac{J_{n}(\nu)}{v(\lambda)^{n+1}}, 
\label{eqn:expdetA}
\end{equation}
produces 
$J_{0}(\nu)=0$, $J_{1}(\nu)={\rm i}\hat{H}_{{\rm long}}$, 
where 
\begin{eqnarray}
\hat{H}_{{\rm long}}&=&-\frac{1}{\sin\nu\cos\nu}
\left[
\sum_{i>j}(\tan\nu)^{i-j}{\rm e}^{2h(\nu)(1-n_{i,-\sigma}-n_{j,-\sigma})}
c_{i\sigma}^{\dagger}c_{j\sigma}
\right.
\nonumber \\
&& \makebox[3em]{}
-\left.
\sum_{i<j}(-\cot\nu)^{i-j}{\rm e}^{-2h(\nu)(1-n_{i,-\sigma}-n_{j,-\sigma})}
c_{i\sigma}^{\dagger}c_{j\sigma}
\right]  \nonumber \\
&& \makebox[1em]{} +U(1+2\sin^{2}\nu)
\sum_{i}\left[
\left(n_{j\uparrow}-\frac{1}{2}\right)
\left(n_{j\downarrow}-\frac{1}{2}\right)
-\frac{1}{4}
\right].
\label{eqn:Hlong}
\end{eqnarray}

Due to the relation (\ref{eqn:DetADetA}), the $J_{n}(\nu)$'s mutually 
commute. Therefore, $\hat{H}_{{\rm long}}$ can 
be embedded in a family of infinite number of commuting operators, 
and can be regarded as a integrable Hamiltonian. 
Moreover, (\ref{eqn:DetAA}) indicates the Y(su(2)) invariance of 
the model;
\begin{equation}
[Q_{0}^{a}(\nu),\hat{H}_{{\rm long}}]=0=
[Q_{1}^{a}(\nu),\hat{H}_{{\rm long}}], \ \ \ (a=1,2,3).
\label{eqn:QH}
\end{equation}
Especially, it implies that the model is su(2) invariant.
By subtracting a constant from $\hat{H}_{{\rm long}}$, we 
can make this Hamiltonian invariant under 
partial particle-hole transformation (\ref{eqn:ph}):
\begin{eqnarray}
\hat{H}_{{\rm long}}^{\prime}&=&-\frac{1}{\sin\nu\cos\nu}
\left[
\sum_{i>j}(\tan\nu)^{i-j}{\rm e}^{2h(\nu)(1-n_{i,-\sigma}-n_{j,-\sigma})}
c_{i\sigma}^{\dagger}c_{j\sigma}
\right.
\nonumber \\
&& \makebox[3em]{}
-\left.
\sum_{i<j}(-\cot\nu)^{i-j}{\rm e}^{-2h(\nu)(1-n_{i,-\sigma}-n_{j,-\sigma})}
c_{j\sigma}^{\dagger}c_{j\sigma}
\right]  \nonumber \\
&& \makebox[1em]{} +U(1+2\sin^{2}\nu)
\sum_{j}
\left(n_{j\uparrow}-\frac{1}{2}\right)
\left(n_{j\downarrow}-\frac{1}{2}\right).
\label{eqn:Hlongp}
\end{eqnarray}
This complicated Hamiltonian can be made simpler by noting (\ref{eqn:U4}) 
to get 
\begin{eqnarray}
\hat{H}_{{\rm long}}^{\prime\prime}
&=& -{\rm i}\sin\nu\cos\nu\hat{H}_{{\rm long}}^{\prime} \nonumber \\
&=&
{\rm i}
\sum_{i>j}\left(\frac{{\rm i}}{r}
\right)^{i-j}{\rm e}^{{\rm i}J(1-n_{i,-\sigma}-n_{j,-\sigma})}
c_{i\sigma}^{\dagger}c_{j\sigma}+\mbox{H.c.}
\nonumber \\
&& \makebox[1em]{} +\frac{6-2r^{2}}{1-r^{2}}\sin J
\sum_{j}
\left(n_{j\uparrow}-\frac{1}{2}\right)
\left(n_{j\downarrow}-\frac{1}{2}\right),
\label{eqn:Hlongpp}
\end{eqnarray}
where $r={\rm i}\cot\nu$, $J=-2{\rm i}h(\nu)$ and $r$ and $J$ are real.
We will, however, use the Hamiltonian  
(\ref{eqn:Hlongp}) instead of (\ref{eqn:Hlongpp}), since it is 
easier to extract informations of the model out of the 
quantum inverse scattering method. 

The Y(su(2)) invariance of $\hat{H}_{{\rm long}}^{\prime}$ 
shown in (\ref{eqn:QH}), together with the invariance under the 
partial particle-hole transformation (\ref{eqn:ph}), leads us to the result
\begin{eqnarray}
&&[Q_{0}^{a}(\nu),\hat{H}'_{{\rm long}}]=0=
[Q_{1}^{a}(\nu),\hat{H}'_{{\rm long}}],\\
&&[Q_{0}^{a\prime}(\nu),\hat{H}'_{{\rm long}}]=0=
[Q_{1}^{a\prime}(\nu),\hat{H}'_{{\rm long}}],
\end{eqnarray}
where $Q_{n}^{a\prime}(\nu)$ is obtained by performing 
the partial particle-hole transformation (\ref{eqn:ph}) to 
$Q_{n}^{a}(\nu)$. 
By construction, the operators $Q_{n}^{a\prime}(\nu)$ form another Y(su(2)) 
algebra, and we can straightforwardly 
verify that $[Q_{m}^{a}(\nu),Q_{n}^{b\prime}(\nu)]$ vanishes for 
$a,b=x,y,z; \ m,n=0,1$.
Putting all things together, we can say that the Hamiltonian 
$\hat{H}'_{{\rm long}}$ is Y(su(2))$\oplus$Y(su(2)) invariant.

Hereafter we shall assume $\hat{H}_{{\rm long}}^{\prime}$ 
to be Hermitian, i.e.
$U$ is real and both $\nu$ and $h(\nu)$ are pure imaginary.
With this assumption, $\hat{H}_{{\rm long}}^{\prime}$
can be regarded as a new Hamiltonian for 
electrons with on-site interaction and 
variable range hopping. The amplitude of the hopping decays exponentially
with the hopping range. There is also an interference effect due 
to the term $\exp(\pm 2h(\nu)(1-n_{i,-\sigma}-n_{j,-\sigma}))$.
One can easily check that in the $\nu=0$ limit the 
hopping terms vanish except for the ones to the neighboring sites, and 
the usual Hubbard model is restored in this limit.
In that sense it is an integrable extension of the Hubbard model 
(\ref{eqn:Hamiltonian})
with variable range hopping.

There is another integrable Hubbard model with variable range hopping 
discovered earlier~\cite{longHubbard}. Its Hamiltonian is given by 
\begin{equation}
H=\sum_{\sigma,i\neq j}t(i-j)c_{i\sigma}^{\dagger}c_{j\sigma}
+U\sum_{j}n_{j\uparrow}n_{j\downarrow}, 
\end{equation}
with 
\begin{equation}
t(s)=-{\rm i}t(-1)^{s}\left( \frac{L}{\pi}\sin\frac{\pi s}{L}\right)^{-1}
\end{equation}
or
\begin{equation}
t(s)=-{\rm i}\sinh\kappa
(-1)^{s}/ \sinh(\kappa s).
\end{equation}
Though both our model and the above model contains the usual 
nearest-neighbor-hopping Hubbard model as a limiting case, 
we do not know whether they can be related with each other.

In closing this section, we shall add a comment.
The Hamiltonian $\hat{H}_{{\rm long}}$ obtained here
is different from $\hat{H}_{\nu}$ obtained from a 
logarithmic derivative of the monodromy matrix ${\cal T}_{mn}(\lambda)_{\nu}$ 
on the finite 
interval. Such things do not occur in the previously studied 
cases of the fermionic nonlinear Schr\"odinger model~\cite{MuWa96b} or the 
Hubbard model~\cite{MuGo97,MuGo98}.
The relation between the two Hamiltonians $\hat{H}_{{\rm long}}$ and 
$\hat{H}_{\nu}$ 
is left as a future problem.

\section{Construction of Eigenvectors}
\setcounter{equation}{0}

\subsection{Creation Operators of Quasiparticles}

As is the case with the usual Hubbard chain and other integrable models, 
the entries of the monodromy matrix ${\tilde{\cal T}}(\lambda)$ 
can be used to construct multiparticle eigenstates on the 
vacuum. By calculating commutators between these entries of 
${\tilde{\cal T}}(\lambda)$ and the particle number operator $\hat{N}$, 
we can see that $B_{a1}(\lambda)_{\nu}$ and $C_{2a}(\lambda)_{\nu}$ add 
$\hat{N}$ by one and $D_{21}(\lambda)_{\nu}$ adds $\hat{N}$ by two, while 
$B_{a2}(\lambda)_{\nu}$ and $C_{1a}(\lambda)_{\nu}$ reduce
$\hat{N}$ by one and $D_{12}(\lambda)_{\nu}$ does by two. 
$A_{ab}(\lambda)_{\nu}$, $D_{11}(\lambda)_{\nu}$ and 
$D_{22}(\lambda)_{\nu}$ keep $\hat{N}$ unchanged. 
To gain further insight, let us calculate
actions of the operators in $\tilde{{\cal T}}(\lambda)$ onto 
the vacuum $|0\rangle$.
By using (\ref{eqn:tildeTsum}), 
some simplest ones are calculated as follows;
\begin{eqnarray}
\label{eqn:B11v}
&&B_{11} (\lambda)_{\nu}
|0\rangle = - \frac{{\rm i}\rho_{6}}{\rho_{9}} 
\sum_{j} 
{\rm e}^{-{\rm i}jp(\lambda,\nu)}
c_{j \downarrow}^{\dagger} |0\rangle , \\
\label{eqn:B21v}
&&B_{21} (\lambda)_{\nu} |0\rangle  =  
\frac{\rho_{6}}{\rho_{9}} 
\sum_{j} 
{\rm e}^{-{\rm i}jp(\lambda,\nu)}
c_{j \uparrow}^{\dagger} |0\rangle , \\
\label{eqn:C21v}
&&C_{21} (\lambda)_{\nu} |0\rangle  =  
- \frac{\rho_{2}}{\rho_{1}} 
\sum_{j} 
{\rm e}^{-{\rm i}jk(\lambda,\nu)}
c_{j \uparrow}^{\dagger} |0\rangle , \\
\label{eqn:C22v}
&&C_{22} (\lambda)_{\nu} |0\rangle  = 
- \frac{{\rm i}\rho_{2}}{\rho_{1}} 
\sum_{j} 
{\rm e}^{-{\rm i}jk(\lambda,\nu)}
c_{j \downarrow}^{\dagger} |0\rangle , \\
\label{eqn:D21v}
&&D_{21} (\lambda)_{\nu} |0\rangle =  
\sum_{j,l}
c_{j\uparrow}^{\dagger}c_{l\downarrow}^{\dagger}
\left[
\theta (j>l)\frac{{\rm i}\rho_{6}\rho_{2}}{\rho_{9}\rho_{1}}
{\rm e}^{-{\rm i}jk(\lambda,\nu)-{\rm i}lp(\lambda,\nu)}
+\theta (j<l)\frac{{\rm i}\rho_{6}\rho_{2}}{\rho_{9}\rho_{1}}
{\rm e}^{-{\rm i}jp(\lambda,\nu)-{\rm i}lk(\lambda,\nu)}
\right.
\nonumber \\
&& \makebox[3em]{}+\left.
\delta_{jl}\frac{{\rm i}\rho_{3}}{\rho_{1}}
{\rm e}^{-{\rm i}j\{ p(\lambda,\nu)+k(\lambda,\nu)\} }
\right]|0\rangle ,
\end{eqnarray}
where $\rho_{j}=\rho_{j}(\lambda,\nu)$ and
\begin{equation}
{\rm e}^{-{\rm i}k(\lambda,\nu)}=
-\rho_{9}(\lambda,\nu)/\rho_{1}(\lambda,\nu), \ \ \ 
{\rm e}^{-{\rm i}p(\lambda,\nu)}=
-\rho_{8}(\lambda,\nu)/\rho_{9}(\lambda,\nu). 
\end{equation}

The commutators between ${\rm Det}_{q}(A(\mu)_{\nu})$ and 
the various operators in 
$\tilde{{\cal T}}(\lambda)_{\nu}$  are 
calculated from the exchange relation (\ref{eqn:newRTT}).  
The resulting commutators are the same as the case of the 
original Hubbard model ($\nu=0$), 
which is summarized in Appendix B.2 of Ref.\cite{MuGo98}.
Hence, the commutators between the Hamiltonian 
$\hat{H}_{{\rm long}}$ and the operators in the matrix 
$\tilde{{\cal T}}(\lambda)_{\nu}$ are also the same as the 
$\nu=0$ case;
\begin{eqnarray}
\label{eqn:HB1}
[\hat{H}_{{\rm long}}, 
B_{a1} (\lambda)_{\nu}] & = & - (2\cos p(\lambda) + U/2) B_{a1} (\lambda)_{\nu}
, \\
\label{eqn:HB2}
[\hat{H}_{{\rm long}}, B_{a2} 
(\lambda)_{\nu}] & = & (2\cos k(\lambda) + U/2) B_{a2} (\lambda)_{\nu}
, \\
\label{eqn:HC1}
[\hat{H}_{{\rm long}}, C_{1a} 
(\lambda)_{\nu}] & = & (2\cos p(\lambda) + U/2) C_{1a} (\lambda)_{\nu}
, \\
\label{eqn:HC2}
[\hat{H}_{{\rm long}}, C_{2a}(\lambda)_{\nu}] & = & - (2\cos k(\lambda) + U/2) C_{2a} (\lambda)_{\nu}
 , \\
\label{eqn:HD1}
[\hat{H}_{{\rm long}}, D_{12} (\lambda)_{\nu}] & = & 
2({\rm e}^{{\rm i}p(\lambda)} + {\rm e}^{- {\rm i}k(\lambda)})
D_{12} (\lambda)_{\nu} , \\
\label{eqn:HD2}
[\hat{H}_{{\rm long}}, D_{21} (\lambda)_{\nu}] & = & 
- 2({\rm e}^{{\rm i}p(\lambda)} + {\rm e}^{- {\rm i}k(\lambda)})
D_{21} (\lambda)_{\nu} ,
\end{eqnarray}
where 
\begin{equation}
{\rm e}^{{\rm i}k(\lambda)}=-{\rm e}^{2h(\lambda)}\cot\lambda, \ \ \ 
{\rm e}^{{\rm i}p(\lambda)}=-{\rm e}^{-2h(\lambda)}\cot\lambda.
\end{equation}
Other entries $D_{11}(\lambda)$, $D_{22}(\lambda)$ and $A_{ab}(\lambda)$ 
commute with $\hat{H}_{{\rm long}}$.
The above results justify the interpretation of 
$B_{a1}(\lambda)$,$C_{2a}(\lambda)$ and $D_{21}(\lambda)$ as creation 
operators. 
$B_{a1}(\lambda)$ and $C_{2a}(\lambda)$ create single particle 
excitations, whereas $D_{21}(\lambda)$ creates a bound state of two 
particles. For example, from $\hat{H}_{{\rm long}}|0\rangle=0$ and 
(\ref{eqn:HC2}) we deduce
\begin{equation}
\hat{H}_{{\rm long}} C_{2a_1} (\lambda_1)_{\nu} \dots C_{2a_n} 
(\lambda_n)_{\nu}
|0\rangle =
- \sum_{j=1}^n (2\cos k(\lambda_j) + U/2) \,
C_{2a_{1}} (\lambda_{1})_{\nu} 
\cdots C_{2a_{n}} (\lambda_{n})_{\nu} |0\rangle .
\end{equation}
Similarly, the applications of the operators $B_{a1}(\lambda)$ or
mixed products of operators $B_{a1}(\lambda)$ and $C_{2a}(\lambda)$ 
on the vacuum produce eigenstates of the Hamiltonian.

Let us consider the relation between these 
creation operators and Yangian Y(su(2)).
Since the Yangian generators $Q_{n}^{a}(\nu)$ ($n = 0, 1$; $a = x, y, z$)  
are coefficients of the power expansion of $A(\mu)_{\nu}$, 
the commutators between $Q_{n}^{a}(\nu)$ and operators 
$B(\lambda)_{\nu}$, $C(\lambda)_{\nu}$, 
and $D(\lambda)_{\nu}$ can be obtained from (\ref{eqn:newRTT}).
The results are the same as in the case of the Hubbard model~\cite{MuGo98};
\begin{eqnarray}
\label{eqn:YB0}
[Q_{0}^{a}(\nu), B (\lambda)_{\nu}]& = & -\frac{1}{2}\tilde{\sigma}^{a} B 
(\lambda)_{\nu} , \\
 \label{eqn:YB1}
[Q_{1}^{a}(\nu), B (\lambda)_{\nu}] & = & \sin p(\lambda)
\tilde{\sigma}^{a} B (\lambda)_{\nu}
+ \frac{U}{2} \varepsilon^{abc}
\tilde{\sigma}^{b} B (\lambda)_{\nu} Q_{0}^{c}(\nu) , \\
\label{eqn:YC0}
[Q_{0}^{a}(\nu), C (\lambda)_{\nu}] & = & \frac{1}{2} C 
(\lambda)_{\nu} \tilde{\sigma}^{a} , \\
\label{eqn:YC1}
[Q_{1}^{a}(\nu), C (\lambda)_{\nu}] & = & - \sin k(\lambda)C
(\lambda)_{\nu}\tilde{\sigma}^{a}
+\frac{U}{2} \varepsilon^{abc}
C (\lambda)_{\nu} \tilde{\sigma}^{b}Q_{0}^{c}(\nu) , \\
\label{eqn:YD}
[Q_{0}^{a}(\nu), D (\lambda)_{\nu}] & = & 
[Q_{1}^{a}(\nu), D (\lambda)_{\nu}] \, = \, 0 .
\end{eqnarray}
These commutators will be used to investigate Yangian 
representations of the multiparticle eigenstates. 

\subsection{Scattering States}

Observing the cases of other integrable models studied 
earlier~\cite{FaTha,PuWuZh87},
we propose the following two pairs of 
normalized creation operators of scattering states, 
\begin{eqnarray}
R_{\alpha}(\lambda)_{\nu}^{\dagger}
&=&{\rm i}^{3-\alpha}\frac{\rho_{1}(\lambda,\nu)}{\rho_{2}(\lambda,\nu)}
\ C_{2\alpha}(\lambda)_{\nu}D_{22}(\lambda)^{-1}_{\nu} \  (\alpha=1,2),
\label{eqn:defRd}\\
\hat{R}_{\alpha}(\lambda)_{\nu}^{\dagger}
&=&{\rm i}^{\alpha-1} \frac{\rho_{9}(\lambda,\nu)}{\rho_{6}(\lambda,\nu)}\ 
B_{3-\alpha,1}(\lambda)_{\nu}D_{11}(\lambda)^{-1}_{\nu} \ (\alpha=1,2).
\label{eqn:defRhd}
\end{eqnarray}
In these formulae 
$\alpha=1$ corresponds to spin-up and $\alpha=2$ 
to spin-down.
The numerical prefactors have been obtained by demanding that 
$R_{\alpha}(\lambda)^{\dagger}_{\nu}$ and 
$\hat{R}_{\alpha}(\lambda)^{\dagger}_{\nu}$ 
generate normalized one-particle states,
\begin{equation}
R_{\alpha}(\lambda)^{\dagger}_{\nu}|0\rangle=\sum_{j}
{\rm e}^{-{\rm i}jk(\lambda,\nu)}
c_{j,\alpha}^{\dagger}|0\rangle, \ \  \
\hat{R}_{\alpha}(\lambda)_{\nu}^{\dagger}
|0\rangle=\sum_{j}
{\rm e}^{-{\rm i}jp(\lambda,\nu)}
c_{j,\alpha}^{\dagger}|0\rangle.
\label{eqn:Roneparticle}
\end{equation}
Hereafter we assume that $\lambda$ is chosen in such a way that 
$R_{\alpha}(\lambda)^{\dagger}_{\nu}$ 
and $\hat{R}_{\alpha}(\lambda)^{\dagger}_{\nu}$ 
create physical states. This means for $R_{\alpha}(\lambda)_{\nu}^{\dagger}$ 
that 
$k(\lambda,\nu)$ 
has to be real and for $\hat{R}_{\alpha}
(\lambda)_{\nu}^{\dagger}$ that 
$p(\lambda,\nu)$ has to be real.

By the method in Ref.~\cite{GoMu97},
hermitian conjugation can be performed to the operators 
$R_{\alpha}(\lambda)_{\nu}$ and $\hat{R}_{\alpha}(\lambda)_{\nu}$, and 
the resulting normalized annihilation operators are 
\begin{eqnarray}
R_{\alpha}(\lambda)_{\nu}
&=&{\rm i}^{2-\alpha}\frac{\rho_{8}(\lambda',\nu)}{\rho_{6}(\lambda',\nu)}
\ D_{11}(\lambda)^{-1}_{\nu}C_{1,3-\alpha}(\lambda)_{\nu},
\label{eqn:defR}\\
\hat{R}_{\alpha}(\lambda)_{\nu}
&=&{\rm i}^{\alpha-2} \frac{\rho_{9}(\lambda',\nu)}{\rho_{2}(\lambda',\nu)}\ 
D_{22}(\lambda)^{-1}_{\nu}B_{\alpha 2}(\lambda)_{\nu},
\label{eqn:defRh}
\end{eqnarray}
with $\lambda'=\pi/2-\lambda^{*}$.
The commutation rules between the operators 
$R_{\alpha}(\lambda)_{\nu}^{\dagger}$,
$\hat{R}_{\alpha}(\lambda)_{\nu}^{\dagger}$,
$R_{\alpha}(\lambda)_{\nu}$,
$\hat{R}_{\alpha}(\lambda)_{\nu}$ are 
\begin{eqnarray}
R_{\alpha}(\lambda)_{\nu}^{\dagger}R_{\beta}(\mu)_{\nu}^{\dagger}
&=&-r(\lambda,\mu)_{\gamma\delta,\alpha\beta}R_{\gamma}(\mu)_{\nu}^{\dagger}
R_{\delta}
(\lambda)_{\nu}^{\dagger},\label{eqn:RdRd}
\\
R_{\alpha}(\lambda)_{\nu}R_{\beta}(\mu)_{\nu}^{\dagger}
&=&-r(\mu,\lambda)_{\gamma\alpha,\delta\beta}R_{\gamma}(\mu)_{\nu}^{\dagger}
R_{\delta}
(\lambda)_{\nu},\label{eqn:RRd}
\\
\hat{R}_{\alpha}(\lambda)_{\nu}^{\dagger}\hat{R}_{\beta}(\mu)_{\nu}^{\dagger}
&=&-r(\mu,\lambda)_{\gamma\delta,\alpha\beta}
\hat{R}_{\gamma}(\mu)_{\nu}^{\dagger}\hat{R}_{\delta}(\lambda)_{\nu}^{\dagger},
\label{eqn:RhdRhd}\\
\hat{R}_{\alpha}(\lambda)_{\nu}\hat{R}_{\beta}(\mu)_{\nu}^{\dagger}
&=&-r(\lambda,\mu)_{\gamma\alpha,\delta\beta}
\hat{R}_{\gamma}(\mu)_{\nu}^{\dagger}\hat{R}_{\delta}(\lambda)_{\nu},
\label{eqn:RhRhd}\\
R_{\alpha}(\lambda)_{\nu}^{\dagger}\hat{R}_{\beta}(\mu)_{\nu}^{\dagger}
&=&-\hat{R}_{\beta}(\mu)_{\nu}^{\dagger}R_{\alpha}(\lambda)_{\nu}^{\dagger}, 
\label{eqn:RdRhd}\\
R_{\alpha}(\lambda)_{\nu}\hat{R}_{\beta}(\mu)_{\nu}^{\dagger}
&=&-\hat{R}_{\beta}(\mu)_{\nu}^{\dagger}R_{\alpha}(\lambda)_{\nu}.
\label{eqn:RRhd}
\end{eqnarray}
Here we neglected $\delta$-function contributions from some 
singular points, e.g., $\lambda=\mu$.

The operators $R_{\alpha}(\lambda)$, $R_{\alpha}(\lambda)^{\dagger}$
and 
$\hat{R}_{\alpha}(\lambda)$, $\hat{R}_{\alpha}(\lambda)^{\dagger}$
form a representation of the graded Zamolodchikov-Faddeev algebra
with $S$-matrix $r(\lambda,\mu)$. These representations may be 
identified as representations of left and right Zamolodchikov-Faddeev 
algebra, respectively~\cite{ZaZa79,FaTha,SmiBOOK,Kha,GoRuSiBOOK}. 
The operators
$R_{\alpha}(\lambda)^{\dagger}$
and $\hat{R}_{\alpha}(\lambda)^{\dagger}$ are graded as odd, 
which implies that they are creation operators 
of fermionic quasi-particles.

We shall present two-particle states generated by 
$R_{\alpha}(\lambda)^{\dagger}$ 
to know physical meanings of $R_{\alpha}^{\dagger}(\lambda)_{\nu}$;
\begin{eqnarray}
&&R_{1}(\lambda)_{\nu}^{\dagger}R_{1}(\mu)_{\nu}^{\dagger}|0\rangle=
\sum_{j,l}c_{j\uparrow}^{\dagger}c_{l\uparrow}^{\dagger}
{\rm e}^{-{\rm i}jk(\lambda,\nu)}{\rm e}^{-{\rm i}lk(\mu,\nu)}|0\rangle, 
\label{eqn:R1R1} \\
&&R_{2}(\lambda)_{\nu}^{\dagger}R_{2}(\mu)_{\nu}^{\dagger}|0\rangle=
\sum_{j,l}c_{j\downarrow}^{\dagger}c_{l\downarrow}^{\dagger}
{\rm e}^{-{\rm i}jk(\lambda,\nu)}{\rm e}^{-{\rm i}lk(\mu,\nu)}|0\rangle, 
\label{eqn:R2R2}\\
&&R_{1}(\lambda)_{\nu}^{\dagger}R_{2}(\mu)_{\nu}^{\dagger}|0\rangle=
\sum_{j,l}c_{j\uparrow}^{\dagger}c_{l\downarrow}^{\dagger}
\left[ \theta(j< l){\rm e}^{-{\rm i}jk(\lambda,\nu)}
{\rm e}^{-{\rm i}lk(\mu,\nu)}
\frac{v(\lambda)-v(\mu)}{v(\lambda)-v(\mu)+{\rm i}U} \right.
\nonumber \\
&&\makebox[2em]{}
+\theta(j< l){\rm e}^{-{\rm i}jk(\lambda,\nu)}
{\rm e}^{-{\rm i}lk(\mu,\nu)}
+\theta(j< l){\rm e}^{-{\rm i}lk(\lambda,\nu)}{\rm e}^{-{\rm i}jk(\mu,\nu)}
\frac{-{\rm i}U}{v(\lambda)-v(\mu)+{\rm i}U} 
\nonumber \\
&&\makebox[2em]{}
+\left. \delta_{jl}
{\rm e}^{-{\rm i}j\{ k(\lambda,\nu)+k(\mu,\nu)\} }
F(\lambda,\mu,\nu)
\right]|0\rangle, 
\label{eqn:R1R2} \\
&&R_{2}(\lambda)_{\nu}^{\dagger}R_{1}(\mu)_{\nu}^{\dagger}|0\rangle=
\sum_{j,l}c_{j\downarrow}^{\dagger}c_{l\uparrow}^{\dagger}
\left[ \theta(j> l){\rm e}^{-{\rm i}jk(\lambda,\nu)}
{\rm e}^{-{\rm i}lk(\mu,\nu)}
\frac{v(\lambda)-v(\mu)}{v(\lambda)-v(\mu)+{\rm i}U} \right.
\nonumber \\
&&\makebox[2em]{}
+\theta(j< l){\rm e}^{-{\rm i}jk(\lambda,\nu)}
{\rm e}^{-{\rm i}lk(\mu,\nu)}
+\theta(j< l){\rm e}^{-{\rm i}lk(\lambda,\nu)}{\rm e}^{-{\rm i}jk(\mu,\nu)}
\frac{-{\rm i}U}{v(\lambda)-v(\mu)+{\rm i}U} 
\nonumber \\
&&\makebox[2em]{}
+\left. \delta_{jl}
{\rm e}^{-{\rm i}j\{ k(\lambda,\nu)+k(\mu,\nu)\} }
F(\lambda,\mu,\nu)
\right]|0\rangle, 
\label{eqn:R2R1}
\end{eqnarray}
where 
\begin{equation}
F(\lambda,\mu,\nu)=\frac{\rho_{9}\rho_{6}}{\rho_{4}\rho_{8}}(\lambda,\mu)
\left(
{\rm e}^{h(\lambda)+h(\mu)-2h(\nu)}\cos\lambda\cos\mu
+{\rm e}^{-h(\lambda)-h(\mu)+2h(\nu)}\sin\lambda\sin\mu
\right).
\end{equation}
From these wavefunctions, like in the case of the Hubbard model, 
we conjecture that 
the $n$-particle state 
\begin{equation}
R_{\alpha_{1}}(\lambda_{1})^{\dagger}_{\nu}\cdots
R_{\alpha_{n}}(\lambda_{n})^{\dagger}_{\nu}|0\rangle
\label{eqn:RR0}
\end{equation}
is a normalized in-state if $k(\lambda_{1},\nu)<\cdots <k(\lambda_{n},\nu)$ 
and
a normalized out-state if $k(\lambda_{1},\nu)>\cdots >k(\lambda_{n},\nu)$.
Here ``normalized'' means that the magnitude of the incident wave is 
unity.
Similarly, for the $\hat{R}$ operators, we conjecture
that
the $n$-particle state 
\begin{equation}
\hat{R}_{\alpha_{1}}(\lambda_{1})^{\dagger}_{\nu}\cdots
\hat{R}_{\alpha_{n}}(\lambda_{n})^{\dagger}_{\nu}|0\rangle
\end{equation}
is a normalized out-state if $p(\lambda_{1},\nu)<\cdots <p(\lambda_{n},\nu)$ 
and
a normalized in-state if $p(\lambda_{1},\nu)>\cdots >p(\lambda_{n},\nu)$.
We have not found out its proof yet.

The Yangian representation of the multiparticle states can be 
investigated by the following commutators from 
(\ref{eqn:YB0})-(\ref{eqn:YD});
\begin{eqnarray}
&&[Q_{0}^{a}(\nu), R_{\alpha}(\lambda)^{\dagger}_{\nu}]=
\frac{1}{2}R_{\beta}(\lambda)^{\dagger}_{\nu}
\sigma_{\beta\alpha}^{a},\label{eqn:Q0R}
\\
&&[Q_{1}^{a}(\nu), R_{\alpha}(\lambda)^{\dagger}_{\nu}]=
-\sin k(\lambda)R_{\beta}(\lambda)^{\dagger}_{\nu}
\sigma_{\beta\alpha}^{a}+\frac{U}{2}\varepsilon^{abc}R_{\beta}
(\lambda)^{\dagger}_{\nu}\sigma_{\beta\alpha}^{b}Q_{0}^{c}(\nu),
\label{eqn:Q1R}
\\
&&[Q_{0}^{a}(\nu), \hat{R}_{\alpha}(\lambda)^{\dagger}_{\nu}]
=\frac{1}{2}\hat{R}_{\beta}
(\lambda)^{\dagger}_{\nu}
\sigma_{\beta\alpha}^{a},
\label{eqn:Q0Rh} 
\\
&&[Q_{1}^{a}(\nu), \hat{R}_{\alpha}(\lambda)^{\dagger}_{\nu}]=
-\sin p(\lambda)\hat{R}_{\beta}
(\lambda)^{\dagger}_{\nu}
\sigma_{\beta\alpha}^{a}-\frac{U}{2}\varepsilon^{abc}\hat{R}_{\beta}
(\lambda)^{\dagger}_{\nu}\sigma_{\beta\alpha}^{b}Q_{0}^{c}(\nu).
\label{eqn:Q1Rh}
\end{eqnarray}
These formulae induce an action of the Yangian on $n$-particle
states~\cite{LeSmi92,MuWa96b,MuGo98}.
Noting that $Q_{0}^{a}(\nu)|0\rangle=0=Q_{1}^{a}(\nu)|0\rangle$, 
we obtain the action of the Yangian on the $n=1$ sector as 
\begin{eqnarray}
Q_{0}^{a}(\nu)R_{\alpha}(\lambda)^{\dagger}_{\nu}|0\rangle&=&
\frac{1}{2}\sigma_{\beta\alpha}^{a}
R_{\beta}(\lambda)^{\dagger}|0\rangle, \\
Q_{1}^{a}(\nu)R_{\alpha}(\lambda)^{\dagger}_{\nu}|0\rangle&=&
-\sin k(\lambda) \sigma_{\beta\alpha}^{a}
R_{\beta}(\lambda)^{\dagger}_{\nu}|0\rangle,
\end{eqnarray}
which is identified as the fundamental representation 
$W_{1}(-2\sin k(\lambda))$.

The $2^{n}$-dimensional representation formed by $n$-particle states  
(\ref{eqn:RR0}) can be studied by the similar manner 
as Ref.~\cite{MuWa96b,MuGo98}, and is identified as 
the tensor product representation 
$W_{1}(-2\sin k(\lambda_{1}))
\otimes \cdots\otimes W_{1}(-2\sin k(\lambda_{n}))$
with co-multiplication $\Delta$ defined by
\begin{eqnarray}
\Delta(Q_{0}^{a})&=&Q_{0}^{a}\otimes 1 + 1 \otimes Q_{0}^{a}, 
\label{eqn:Delta0}
\\
\Delta(Q_{1}^{a})&=&Q_{1}^{a}\otimes 1 + 1 \otimes Q_{1}^{a}
+U\varepsilon^{abc}Q_{0}^{b}\otimes Q_{0}^{c}.
\label{eqn:Delta1}
\end{eqnarray}
This representation is irreducible since $k(\lambda_{i})$'s are real.
Hence, we conclude that all the $n$-particle states of (\ref{eqn:RR0}) 
can be constructed by applying the Yangian generators 
$Q_{n}^{a}(\nu)$ to the highest weight state
\begin{equation}
R_{\uparrow}(\lambda_{1})^{\dagger}_{\nu}\cdots
R_{\uparrow}(\lambda_{n})^{\dagger}_{\nu}|0\rangle,
\end{equation}
which is clearly proportional to 
\begin{equation}
c_{\uparrow}(k(\lambda_{1},\nu))^{\dagger}\cdots
c_{\uparrow}(k(\lambda_{n},\nu))^{\dagger}|0\rangle,
\end{equation}
with $c_{\alpha}(k)^{\dagger}=\sum_{j}{\rm e}^{-{\rm i}jk}
c_{j\alpha}^{\dagger}$. 
If we use $\hat{R}$ operators instead of $R$, we reach the 
similar conclusion with $k(\lambda)$ replaced by $p(\lambda)$ and 
the definition of the comultiplication changed to 
\begin{eqnarray}
\Delta'(Q_{0}^{a})&=&Q_{0}^{a}\otimes 1 + 1 \otimes Q_{0}^{a}, 
\label{eqn:Delta'0}
\\
\Delta'(Q_{1}^{a})&=&Q_{1}^{a}\otimes 1 + 1 \otimes Q_{1}^{a}
-U\varepsilon^{abc}Q_{0}^{b}\otimes Q_{0}^{c}.
\label{eqn:Delta'1}
\end{eqnarray}

\subsection{Bound States}

In order to investigate structures of bound states, let us begin with the 
two-particle bound states. Among two-particle states 
\begin{equation}
C_{2a}(\lambda_{1})_{\nu}C_{2b}(\lambda_{2})_{\nu}|0\rangle,
\label{eqn:C2aC2b}
\end{equation}
we should set $(a,b)=(2,1),(1,2)$ in order to obtain bound states, 
which follows from explicit calculation of wavefunctions.
In the former case, 
in which the eigenstate is calculated as
\begin{eqnarray}
&&C_{22}(\lambda_{1})_{\nu}C_{21}(\lambda_{2})_{\nu}|0\rangle\propto
\sum_{j,l}c_{j\downarrow}^{\dagger}c_{l\uparrow}^{\dagger}
\left[ \theta(j> l)
(v(\lambda_{1})-v(\lambda_{2})){\rm e}^{-{\rm i}jk(\lambda_{1},\nu)}
{\rm e}^{-{\rm i}lk(\lambda_{2},\nu)}
 \right.
\nonumber \\
&&\makebox[1em]{}
+\theta(j< l)
(v(\lambda_{1})-v(\lambda_{2})+{\rm i}U)
{\rm e}^{-{\rm i}jk(\lambda_{1},\nu)}{\rm e}^{-{\rm i}lk(\lambda_{2},\nu)}
+\theta(j< l)(-{\rm i}U)
{\rm e}^{-{\rm i}lk(\lambda_{1},\nu)}{\rm e}^{-{\rm i}jk(\lambda_{2},\nu)} 
\nonumber \\
&&\makebox[1em]{}
+\left. \delta_{jl}
{\rm e}^{-{\rm i}j\{ k(\lambda_{1},\nu)+k(\lambda_{2},\nu)\} }
(v(\lambda_{1})-v(\lambda_{2})+{\rm i}U)F(\lambda_{1},\lambda_{2},\nu)
\right]|0\rangle. 
\label{eqn:C2sC21}
\end{eqnarray}
The condition for it to be a bound state is 
\begin{eqnarray}
&&v(\lambda_{1})-v(\lambda_{2})=-{\rm i}U, 
\label{eqn:cond1}
\\
&&\Im k(\lambda_{1},\nu)=-\Im k(\lambda_{2},\nu)<0.
\label{eqn:cond2}
\end{eqnarray}
Provided that these conditions hold, it follows that
\begin{equation}
C_{22}(\lambda_{1})_{\nu}C_{21}(\lambda_{2})_{\nu}
=-C_{21}(\lambda_{1})_{\nu}C_{22}(\lambda_{2})_{\nu},
\end{equation}
which implies that $(a,b)=(2,1)$ and $(2,1)$ cases give the same bound state. 
To summarize, among the two-particle states (\ref{eqn:C2aC2b}),
there is only one bound state, which is achieved in the case 
$(a,b)=(2,1)$ with the conditions (\ref{eqn:cond1}),(\ref{eqn:cond2}).
We have not, however, succeeded to investigate 
this condition further due to the complicated form of the 
function $k(\lambda,\nu)$. 

Due to this complicated form of $k(\lambda,\nu)$, 
we have not found out general forms of multiparticle bound states 
or their creation operators.
But if we simply mimic the construction of bound state operators of the 
original Hubbard model in Ref.~\cite{MuGo98}, 
we can formally make ``bound state operators'' by 
\begin{eqnarray}
&&C^{(2m)}_{2}(\lambda_{1},\cdots,\lambda_{2m})_{\nu}=
C_{22}(\lambda_{1})_{\nu}C_{21}(\lambda_{2})_{\nu}
C_{22}(\lambda_{3})_{\nu}C_{21}(\lambda_{4})_{\nu}\cdots
C_{22}(\lambda_{2m-1})_{\nu}C_{21}(\lambda_{2m})_{\nu}, \makebox[1em]{} \\
&&D^{(2m)}_{22}(\lambda_{1},\cdots,\lambda_{2m})_{\nu}=
D_{22}(\lambda_{1})_{\nu} D_{22}(\lambda_{2})_{\nu}
\cdots D_{22}(\lambda_{2m-1})_{\nu}D_{22}(\lambda_{2m})_{\nu}. \\
&&  R^{(2m)}(\lambda_{1},\cdots,\lambda_{2m})_{\nu}^{\dagger}=
C^{(2m)}_{2}(\lambda_{1},\cdots,\lambda_{2m})_{\nu}
D^{(2m)}_{22}(\lambda_{1},\cdots,\lambda_{2m})_{\nu}^{-1},
\end{eqnarray}
where
\begin{eqnarray}
&& k(\lambda_{2s})+p(\lambda_{2s-1})=\pi  \ ({\rm mod} \  2\pi),\\
&&\sin k(\lambda_{2s-1})=\sin k(\lambda_{1})+\frac{{\rm i}U(s-1)}{2},
\ (s=1,\cdots,m).
\end{eqnarray}
The commutation rules of $R^{(2m)\dagger}$
with $R^{(2n)\dagger}$, $R^{\dagger}$ or
$Q_{n}^{a}(\nu)$ are the same as those in the case of the usual 
Hubbard model ((6.58)-(6.60) in Ref.~\cite{MuGo98}).
But the serious problem with this operator $R^{(2m)\dagger}$ is that 
we do not know whether it certainly produces physical states.
As the case of two-particle bound states ($m=1$) is already 
difficult to study, 
there is little hope that we can 
get deep understanding of multiparticle bound state operators.

\section{Concluding Remarks and Discussion}
\setcounter{equation}{0}

In this paper we have introduced new integrable 
variant of the nearest-neighbor Hubbard model 
with variable range hopping. We have constructed it 
by the quantum inverse scattering method on the 
infinite interval at zero density, using the one-parameter deformation of 
the ${\cal L}$-matrix of the Hubbard model.
By construction, together with the knowledge of the case 
of the Hubbard model studied earlier, 
this Hamiltonian is among an infinite number of commuting operators 
and thus integrable. Moreover, it commutes with operators 
$Q_{n}^{a}(\nu)$ $(n=0,1; \ a=x,y,z)$, which form a representation of 
the Y(su(2)) Yangian. If we take the 
Hamiltonian $\hat{H}_{{\rm long}}^{\prime}$ or 
$\hat{H}_{{\rm long}}^{\prime\prime}$ instead of $\hat{H}_{{\rm long}}$, 
it is invariant under the partial particle-hole transformation and 
is Y(su(2))$\oplus$Y(su(2)) invariant.
Normalized creation and annihilation operators of quasiparticles are 
explicitly constructed and are shown to 
form the Zamolodchikov-Faddeev algebra. 
Multiparticle scattering states are constructed with these operators, 
while bound states still require further study.

The merits of deriving the Hamiltonian along such line are as follows.
First, the existence of the Yangian symmetry can be established without 
any ad hoc methods. Second, the forms of multiparticle states upon the 
zero-density vacuum can be derived without any ansatz.
They are calculated directly from actions of the elements 
of the monodromy matrix $\tilde{{\cal T}}(\lambda)_{\nu}$ on the vacuum.
Derivation of multiparticle wavefunctions by making some kind of 
``Bethe ansatz" 
is rather difficult due to the complicated structure of the Hamiltonian.

Although it is not easy to interpret the physical meaning of the term 
of the phase factor 
$\exp(\pm 2h(\nu)(1-n_{i,-\sigma}-n_{j,-\sigma}))$ 
in the Hamiltonian, 
it would be interesting to investigate 
thermodynamic properties of this new integrable model and 
it will be studied in separate papers~\cite{Mu98}.

\section*{Acknowledgements}
The author 
benefited from stimulating discussions with F.~G\"ohmann and   
M.~Shiroishi. 
The author is also grateful to M.~Wadati and N.~Nagaosa for their 
continuous encouragements.

\setcounter{section}{0}
\renewcommand{\thesection}{Appendix \Alph{section}}

\section{Expressions of the $\check{{\cal R}}$-matrix and the %
${\cal L}$-matrix}
\label{appendix:RL}
\renewcommand{\theequation}{\Alph{section}.\arabic{equation}}
\setcounter{equation}{0}

In this appendix we present the expressions of the 
$\check{{\cal R}}$-matrix and the ${\cal L}$-matrix. The 
${\cal L}$-matrix~\cite{OlWaAk87} is given as
\begin{equation}
{\cal L}_{j}(\lambda)={\rm e}^{h(\lambda)\sigma^{z}\otimes\sigma^{z}/2}
\left[
\left(
\begin{array}{cc}
-f_{j\uparrow}(\lambda)&{\rm i}c_{j\uparrow} \\
c_{j\uparrow}^{\dagger} & g_{j\uparrow}(\lambda)
\end{array}
\right)
\otimes_{s}
\left(
\begin{array}{cc}
f_{j\downarrow}(\lambda)& c_{j\downarrow} \\
{\rm i}c_{j\downarrow}^{\dagger} & -g_{j\downarrow}(\lambda)
\end{array}
\right)
\right]
{\rm e}^{h(\lambda)\sigma^{z}\otimes\sigma^{z}/2}
\end{equation}
with the grading $P(1)=0$, $P(2)=1$ and functions
\begin{equation}
f_{j\sigma}(\lambda)=({\rm i}\cot\lambda )^{n_{j\sigma}}\sin\lambda, \ \ \ 
g_{j\sigma}(\lambda)=(-{\rm i}\tan\lambda )^{n_{j\sigma}}\cos\lambda.
\end{equation}
The $\check{{\cal R}}$-matrix is given as~\cite{OlWaAk87} 
\begin{eqnarray}
\lefteqn{{\check {\cal R}}(\lambda,\mu) 
= }\nonumber \\[1ex]
& & \hspace{-22pt} \left( 
{\arraycolsep 2pt
\begin{array}{cccccccccccccccc}
\rho_{1} & 0 & 0 & 0 & 
0 & 0 & 0 & 0 & 0 & 0 & 0 & 0 & 0 & 0 & 0 & 0 \\
0 & \rho_{2} & 0 & 0 & 
{\rm i}\rho_{9} 
& 0 & 0 & 0 & 0 & 0 & 0 & 0 
& 0 & 0 & 0 & 0 \\
0 & 0 & \rho_{2} & 0 & 
0 & 0 & 0 & 0 &
{\rm i}\rho_{9} 
& 0 & 0 & 0 & 0 & 0 & 0 & 0 \\
0 & 0 & 0 & \rho_{3} & 
0 & 0 & -{\rm i}\rho_{6} & 0 &
0 & {\rm i}\rho_{6} & 0 & 0 &
-\rho_{8} & 0 & 0 & 0 \\
0 & 
 -{\rm i}\rho_{10} &0 & 0 & \rho_{2} &  0 & 0 & 0 &
0 & 0 & 0 & 0 &
0 & 0 & 0 & 0 \\
0 & 0 & 0 & 0 & 
0 &  \rho_{4} & 0 & 0 &
0 & 0 & 0 & 0 &
0 & 0 & 0 & 0 \\
0 & 0 & 0 & {\rm i}\rho_{6} & 
0 & 0 & \rho_{5} & 0 & 0 & 
-\rho_{7} & 0 & 0 &
-{\rm i}\rho_{6} & 0 & 0 & 0 \\
0 & 0 & 0 & 0 & 
0 & 0 & 0 & \rho_{2} &
0 & 0 & 0 & 0 &
0 & 
-{\rm i}\rho_{10}
 & 0 & 0 \\
0 & 0 &  -{\rm i}\rho_{10} & 0 & 
0 & 0 & 0 & 0 &
\rho_{2} & 0 & 0 & 0 &
0 & 0 & 0 & 0 \\
0 & 0 & 0 & -{\rm i}\rho_{6} & 
0 & 0 & -\rho_{7}
& 0 & 0 &  \rho_{5}
 & 0 & 0 & 
{\rm i}\rho_{6} & 0 & 0 & 0 \\
0 & 0 & 0 & 0 & 
0 & 0 & 0 & 0 &
0 & 0 &  \rho_{4} & 0 &
0 & 0 & 0 & 0 \\
0 & 0 & 0 & 0 & 
0 & 0 & 0 & 0 &
0 & 0 & 0 & \rho_{2}&
0 & 0 & -{\rm i}\rho_{10}
& 0 \\
0 & 0 & 0 &  -\rho_{8} & 
0 & 0 & {\rm i}\rho_{6} & 0 &
0 & -{\rm i}\rho_{6} & 0 & 0 &
\rho_{3}& 0 & 0 & 0 \\
0 & 0 & 0 & 0 & 
0 & 0 & 0 &  {\rm i} \rho_{9} &
0 & 0 & 0 & 0 &
0 & \rho_{2} & 0 & 0 \\
0 & 0 & 0 & 0 & 
0 & 0 & 0 & 0 &
0 & 0 & 0 &  {\rm i} \rho_{9} &
0 & 0 & \rho_{2} & 0 \\
0 & 0 & 0 & 0 & 
0 & 0 & 0 & 0 &
0 & 0 & 0 & 0 &
0 & 0 & 0 &  \rho_{1} 
\end{array}
}
\right)
\raisebox{-22ex}{,} \makebox[2em]{}
\end{eqnarray}
where $\rho_{j}=\rho_{j}(\lambda,\mu)$ is defined by
\begin{eqnarray*}
\rho_{1}&=&({\rm e}^{l}\cos\lambda\cos\mu+{\rm e}^{-l}\sin\lambda\sin\mu)
\rho_{2}, \\ 
\rho_{4}&=&({\rm e}^{l}\sin\lambda\sin\mu+{\rm e}^{-l}\cos\lambda\cos\mu)
\rho_{2}, \\ 
\rho_{9}&=&(-{\rm e}^{l}\cos\lambda\sin\mu+{\rm e}^{-l}\sin\lambda\cos\mu)
\rho_{2}, \\ 
\rho_{10}&=&({\rm e}^{l}\sin\lambda\cos\mu-{\rm e}^{-l}\cos\lambda\sin\mu)
\rho_{2}, \\ 
\rho_{3}&=&\frac{
{\rm e}^{l}\cos\lambda\cos\mu-{\rm e}^{-l}\sin\lambda\sin\mu}{
\cos^{2}\lambda-\sin^{2}\mu}
\rho_{2}, \\ 
\rho_{5}&=&\frac{
-{\rm e}^{l}\sin\lambda\sin\mu+{\rm e}^{-l}\cos\lambda\cos\mu}{
\cos^{2}\lambda-\sin^{2}\mu}
\rho_{2}, \\ 
\rho_{6}&=&\frac{
{\rm e}^{-2h(\mu)}\cos\lambda\sin\lambda-{\rm e}^{-2h(\lambda)}
\cos\mu\sin\mu}{
\cos^{2}\lambda-\sin^{2}\mu}
\rho_{2}, \\ 
\rho_{7}&=&\rho_{5}-\rho_{4}, \ \ \ \  
\rho_{8}=\rho_{3}-\rho_{1},
\end{eqnarray*}
with $l=h(\lambda)-h(\mu)$.
The function $h(\lambda)$ is defined by (\ref{eqn:U4}).
These matrices 
are identical with those in Ref.~\cite{OlWaAk87}, except for the point 
that the spectral parameters $\lambda$ and $\mu$ are shifted by $\pi/4$.
These matrices satisfy the exchange relation (\ref{eqn:RLL}).
Its one-parameter deformed version is given by (\ref{eqn:checkRLLnu}) 
with the $\check{{\cal R}}$-matrix presented above and the new 
${\cal L}$-matrix given by 
\begin{eqnarray*}
&&{\cal L}_{j}(\lambda)_{\nu}^{11}=
\rho_{1}n_{j\uparrow}n_{j\downarrow}
-{\rm i}\rho_{10}(n_{j\uparrow}+n_{j\downarrow}-2n_{j\uparrow}n_{j\downarrow})
-\rho_{8}(1-n_{j\uparrow})(1-n_{j\downarrow}),
\\
&&{\cal L}_{j}(\lambda)_{\nu}^{12}=-{\rm i}\rho_{2}n_{j\uparrow}c_{j\downarrow}
-\rho_{6}(1-n_{j\uparrow})c_{j\downarrow},\\
&&{\cal L}_{j}(\lambda)_{\nu}^{13}=-\rho_{2}c_{j\uparrow}n_{j\downarrow}
+{\rm i}\rho_{6}c_{j\uparrow}(1-n_{j\downarrow}),\\
&&{\cal L}_{j}(\lambda)_{\nu}^{14}={\rm i}\rho_{3}c_{j\uparrow}c_{j\downarrow}
,\\
&&{\cal L}_{j}(\lambda)_{\nu}^{21}=
\rho_{2}n_{j\uparrow}c^{\dagger}_{j\downarrow}
-{\rm i}\rho_{6}(1-n_{j\uparrow})c^{\dagger}_{j\downarrow},\\
&&{\cal L}_{j}(\lambda)_{\nu}^{22}=
\rho_{9}n_{j\uparrow}n_{j\downarrow}
+{\rm i}\rho_{4}n_{j\uparrow}(1-n_{j\downarrow})
-{\rm i}\rho_{7}(1-n_{j\uparrow})n_{j\downarrow}
+\rho_{9}(1-n_{j\uparrow})(1-n_{j\downarrow}),
\\
&&{\cal L}_{j}(\lambda)_{\nu}^{23}=
\rho_{5}c_{j\uparrow}c^{\dagger}_{j\downarrow}
,\\
&&{\cal L}_{j}(\lambda)_{\nu}^{24}=\rho_{6}c_{j\uparrow}n_{j\downarrow}
+{\rm i}\rho_{2}c_{j\uparrow}(1-n_{j\downarrow}),\\
&&{\cal L}_{j}(\lambda)_{\nu}^{31}={\rm i}\rho_{2}c^{\dagger}_{j\uparrow}
n_{j\downarrow}
+\rho_{6}c^{\dagger}_{j\uparrow}(1-n_{j\downarrow}),\\
&&{\cal L}_{j}(\lambda)_{\nu}^{32}=
\rho_{5}c^{\dagger}_{j\uparrow}c_{j\downarrow}
,\\
&&{\cal L}_{j}(\lambda)_{\nu}^{33}=
\rho_{9}n_{j\uparrow}n_{j\downarrow}
-{\rm i}\rho_{7}n_{j\uparrow}(1-n_{j\downarrow})
+{\rm i}\rho_{4}(1-n_{j\uparrow})n_{j\downarrow}
+\rho_{9}(1-n_{j\uparrow})(1-n_{j\downarrow}),
\\
&&{\cal L}_{j}(\lambda)_{\nu}^{34}=-{\rm i}\rho_{6}n_{j\uparrow}c_{j\downarrow}
+\rho_{2}(1-n_{j\uparrow})c_{j\downarrow},\\
&&{\cal L}_{j}(\lambda)_{\nu}^{41}=-{\rm i}\rho_{3}c^{\dagger}_{j\uparrow}
c^{\dagger}_{j\downarrow}
,\\
&&{\cal L}_{j}(\lambda)_{\nu}^{42}=-{\rm i}\rho_{6}c^{\dagger}_{j\uparrow}
n_{j\downarrow}
+\rho_{2}c^{\dagger}_{j\uparrow}(1-n_{j\downarrow}),\\
&&{\cal L}_{j}(\lambda)_{\nu}^{43}=
\rho_{6}n_{j\uparrow}c^{\dagger}_{j\downarrow}
+{\rm i}\rho_{2}(1-n_{j\uparrow})c^{\dagger}_{j\downarrow},\\
&&{\cal L}_{j}(\lambda)_{\nu}^{44}=
\rho_{8}n_{j\uparrow}n_{j\downarrow}
+{\rm i}\rho_{10}(n_{j\uparrow}+n_{j\downarrow}-2n_{j\uparrow}n_{j\downarrow})
-\rho_{1}(1-n_{j\uparrow})(1-n_{j\downarrow}),
\end{eqnarray*}
where $\rho_{j}=\rho_{j}(\lambda,\nu)$.
One can easily check that ${\cal L}_{j}(\lambda)_{\nu=0}
\propto {\cal L}_{j}(\lambda)$.

\end{document}